\documentclass{aa}

\usepackage{graphicx}
\usepackage{txfonts}
\usepackage{color}
\usepackage{multirow}
\usepackage{hyperref}
\usepackage{ulem}
\hypersetup{colorlinks = true, citecolor = {blue}}

\begin{document}
\title{Merging timescale for supermassive black hole binary in interacting galaxy NGC~6240}
\subtitle{}

\author{M. Sobolenko\inst{1}\fnmsep\thanks{ORCID:
        https://orcid.org/0000-0003-0553-7301}
        \and
        P. Berczik\inst{2,3,1}\fnmsep\thanks{ORCID: https://orcid.org/0000-0003-4176-152X}
        \and
        R. Spurzem\inst{2,3,4}\fnmsep\thanks{Research Fellow at Frankfurt Institute for Advanced Studies (FIAS)}
        }

\institute{Main Astronomical Observatory,
           National Academy of Sciences of Ukraine,
           27 Akademika Zabolotnoho St., 03143 Kyiv, Ukraine,
           \email{sobolenko@mao.kiev.ua}
           \and
           National Astronomical Observatories and Key Laboratory of Computational Astrophysics,
					 Chinese Academy of Sciences, 20A Datun Rd., Chaoyang District,
					 100101 Beijing, China
           \and
     	   Astronomisches Rechen-Institut,
     	              Zentrum f\"ur Astronomie, University of Heidelberg,
                      M\"onchhofstrasse 12-14, 69120 Heidelberg, Germany
           \and
           Kavli Institute for Astronomy and Astrophysics,
           Peking University,
           Yiheyuan Lu 5, Haidian Qu, 100871 Beijing, China
           }

\date{Received XXX; accepted YYYY}

\abstract
{One of the main possible way of creating the supermassive black hole (SMBH) is a so call
hierarchical merging scenario. Central SMBHs at the final phase of interacting and
coalescing host-galaxies are observed as SMBH binary (SMBHB) candidates at different
separations from hundreds of pc to mpc.}
{Today one of the strongest SMBHB candidate is a ULIRG galaxy NGC~6240 which was X-ray
spatially and spectroscopically resolved by Chandra. Dynamical calculation of central
SMBHBs merging in a dense stellar environment allows us to retrace their evolution
from kpc to mpc scales. The main goal of our dynamical modeling was to reach the
final, gravitational wave (GW) emission regime for the model BHs.}
{We present the direct N-body simulations with up to one million particles and relativistic
post-Newtonian corrections for the SMBHs particles up to 3.5 post-Newtonian terms.}
{Generally speaking, the set of initial physical conditions
can strongly effect of our merging time estimations. But in a range of our parameters,
we did not find any strong correlation between merging time
and BHs mass or BH to bulge mass ratios. Varying the model
numerical parameters (such as particle number - N) makes our results quite robust and
physically more motivated. From our 20 models we found the upper limit of merging time for
central SMBHB is less than $\sim$55 Myr. This concrete number are valid only for our set of
combination of initial mass ratios.}
{Further detailed research of rare dual/multiple BHs in dense stellar environment
(based on observations data) can clarify the dynamical co-evolution of central BHs and their
host-galaxies.}


\titlerunning{Merging time for SMBHB at NGC~6240}
\authorrunning{M. Sobolenko et al.}
\maketitle

\section{Introduction} \label{sec:intro}

Majority of classical bulges and elliptical galaxies are harbors for the central
supermassive black holes (SMBHs) and even some part (<20\%) of dwarf galaxies
probably can contain them \citep{Kormendy2013, Mezcua2017}. SMBH with mass
M = 4.28$\times10^6$ M$_{\odot}$ \citep{Gillessen2017} at the center of Milky Way
is most powerful evidence of correctness of this statement. Theoretically SMBHs
can grow to the observable masses (10$^6$ -- 10$^{10}$ M$_{\odot}$) in different ways.
One of the possible mechanisms for the SMBHs formation is through a hierarchical
merging of host-galaxies, according to the standard $\Lambda$CDM cosmology
\citep{White1978}. Co-evolution of SMBHs and galaxies manifests via strong
correlations between the SMBH mass and galaxy bulges parameters (velocity dispersion,
bulge luminosity/mass, etc.) \citep{Kormendy2013} or also over the strong correlation
between the SMBH's mass and global star formation properties in the host galaxies
\citep{2020ApJ...897..102C}.

From observations we expect to find a SMBH's at different merging stages with a
strong gravitational waves emission detection at the last phase of coalescence.
Observations at nanohertz frequencies ($\sim$1--100~nHz) accesible to Pulsar
Timing Arrays (PTAs) \citep{Sesana2008,Burke2019} predict detection non-oscillatory
components of gravitational wave (GW) signals for SMBH binary (SMBHB) coalescence
at the high end of mass distribution in the the next 5--10~years \citep{Taylor2019}.
GW from Super-massive (<10$^7$ M$_{\odot}$) and Intermediate mass BH's could
be directly detected by the Laser Interferometer Space Antenna (LISA) space
mission \citep{LISA2013} already in our decade.

Dynamical evolution of the SMBHB in dense stellar environment (gas-free, so called "dry", merging)
consists of three main parts \citep{Begelman1980}. At first stage central BHs surrounded by stars
approaching via efficient dynamical friction.
Assuming the galaxy center density distribution as a singular isothermal sphere, central BHs each
with masses $m_{\bullet}$ are sinking toward the center of the stellar distribution, contains $N$ stars,
with the density $\rho=\sigma^{2}/(2\pi Gr_{c}^{2})$ at the timescale:
\begin{equation}
\tau_{\rm df} \sim 2\times10^{8}{\frac{1}{{\rm ln}(N)}}~\Bigl(\frac{10^{6}\; \rm M_{\odot}}{m_{\bullet}}\Bigr) \Bigl(\frac{r_{c}}{100 \;\rm pc}\Bigr)^{2} \Bigl(\frac{\sigma}{100 \;\rm km \; s^{-1}}\Bigr) \; \rm [yr],
\end{equation}
where $r_{c}$ is a core radius, $\sigma$ is a velocity dispersion and $G$ is the
gravitational constant \citep{Yu2002}. BHs at this stage are effectively losing
energy and form bound pair with semimajor axis $a_{\rm b}$ which nearly equals the
BH's influence radius.

SMBHs continue to in-spiral to each other due to decreasing dynamical friction. By definition,
the influence radius is a radius which contains as much as twice the mass of the central
SMBHB $M_{\bullet}=m_{\bullet1}+m_{\bullet2}$, where primary BH mass $m_{\bullet1}$ is
bigger then seconadry BH mass $m_{\bullet2}$. Furthermore three-body interactions
(scattering) with passing stars becomes dominant and the second phase starts when
the binary semimajor axis reaches the values $a_{\rm h}\equiv G\mu/4\sigma^{2}$,
where $\mu=m_{\bullet 1}m_{\bullet 2}/(m_{\bullet 1}+m_{\bullet 2})$ is a reduced
mass \citep{Quinlan1996,Yu2002}. Binary with semimajor axis $a$ is effectively hardening
at timescale:
\begin{equation}
\tau_{\rm hard} \sim 70~\Bigl(\frac{\sigma}{100\; \rm km \;s^{-1}}\Bigr)\Bigl(\frac{10^{4}\rm\; M_{\odot} \;pc^{-3}}{\rho}\Bigr) \Bigl(\frac{10^{-3} \rm \; pc}{a}\Bigr) \; \rm [Myr].
\end{equation}

At the final stage the GW radiation can effectively carry out the binary kinetic energy
and angular momentum. Assuming the simple post-Newtonian ($\mathcal{PN}$) formalism,
and using only the 2.5$\mathcal{PN}$ corrections, the GW merging timescale for the binary
with eccentricity $e$ can be described as:
\begin{equation}
\tau_{\rm gw} = \frac{5}{64} \frac{c^{5}a^{4}}{G^{3}\mu M_{\bullet}^{2}}\frac{(1-e^{2})^{7/2}}{1+73e^{2}/24+37e^{4}/96}\;\rm [s]
\end{equation}
where $c$ is a speed of the light and $M_{\bullet}$ is the total BHs mass
\citep{Peters1963,Peters1964a,Peters1964b}. At final merging stage the binary
lifetime rapidly decreases and the orbits tend to be circularize. SMBHs final
merging can also have a velocity "recoil" with the speed up to $5000$~km~s$^{-1}$
\citep{Peres1962,Bekenstejn1973,Lousto2019}.

The above described scenario and timescales just show the principal evolutionary path. The timescale for
transition from one phase to another is not a well fixed value. As an example, in a case of extreme mass
ratio $m_{\bullet2}/m_{\bullet1}\ll1$ the GW gravitational radiation can start dominating in dissipation
before the strong hardening phase is begins \citep{Yu2002}.
Also a well known "final parsec problem" can even stalls the SMBHB orbital evolution and significantly increase the
coalescence time to more than a Hubble time \citep{Milosavljevic2001,Milosavljevic2003}. In a gas-free ideal stellar
spherical system a depletion of the loss-cone is quite possible. When there are not enough stars for the energy and
angular momentum effective carrying out. But numerical simulations applying a triaxial geometry based on the observed
merging remnants with the sufficient rate of non-axisymmetry already shows a not stalling BHs hardening
\citep{Berczik2006,Gualandris2017}.

Presence of the gas (gas-rich, so called "wet", merging) influence to the timescales of different merging phases
and can give one more solution for the "final parsec problem". \citep{Mayer2007} firstly showed the evolution of
BHs in gas-rich merging from tens kpc to several pc in around 5 Gyr. Variety of researches showed the complex and
different role of gas in minor (mass ratio$\leq$10) and major merging (mass ratio$>$10): gas can both accelerate
shrinking of binary orbit and delay its evolution. In major merging gas can help in binary pairing and hardening
in formed post-merged gas disk \citep{Escala2005,Cuadra2009,SouzaLima2020} and via interaction of gas clumps with
binary \citep{Goicovic2017,Goicovic2018}. The role of gas in minor merging is strongly depend on initial geometry
and gas content \citep{Callegari2009,Callegari2011}. For heavy SMBHB (with BHs mass ratio $\sim$1 and masses more
than $10^{7}$M$_{\odot}$) gas disks affect the merging processes at significantly lower rate \citep{Cuadra2009,Lodato2009}.
Just a one simulation \citep{Khan2016} showed \textit{ab-initio} SMBHB evolution starting from cosmological
simulation to GW emission.

As mentioned before, we should observe SMBHs coalescence at different merging stages: as double active galactic
nuclei (AGN, at separations from tens of kpc to kpc), SMBHB (at separations from pc to sub-pc scale) and recoiling BH.
The interacting galaxy NGC~6240 was a firstly directly detected (X-ray spectroscopic and image confirmation) and spatially
resolved double AGN candidate \citep{Komossa2003}. Strong double AGN candidates (with separations $\Delta R<1$ kpc, are the most
suitable for N-body simulations) was serendipitously discovered: 4C +37.11 \citep{Maness2004,Rodriguez2006},
SDSS J115822+323102 \citep{Muller2015}, NGC3393 \citep{Fabbiano2011}, IC4553 \citep{Paggi2013},
Mrk 273 \citep{UVivian2013}, SDSS J132323-015941 \citep{Woo2014}. Systematic search gave samples of double AGN
candidates at optic \citep{McGurk2011, Fu2012, McGurk2015, Ellison2017}, X-ray (usually previously selected from
another band \citep{Green2011, Gross2019}), infrared \citep{Satyapal2017, Pfeifle2019} and radio
\citep{Burke-Spolaor2011} bands each of which requires more detail cross-bands studying. Observing of spatial
offset SMBHs from the host galaxy stellar center or broad emission line velocity offset from systematic velocity
gave us a dozen candidates to recoiling SMBHs. First such a candidate was a luminous quasar SDSS J0927+2943 with
offset by $\sim 2650\rm\;km\;s^{-1}$ \citep{Komossa2008}. In our paper, for simplification, we will use the term
SMBHB regardless of the component separation.

We structure the paper in following way. We describe the physical parameters of the galaxy NGC~6240 and central BHs
which we use for N-body simulations in Section~\ref{sec:phys_param}. How we construct the physical and numerical
N-body models are detailed in Section~\ref{sec:ini_param}. Our code is presented in Section~\ref{sec:code}.
The motion of BHs as relativistic particles and their hardening are represented using post-Newtonian formalism
in Section~\ref{sec:PN_hard}. The description of merging BHs evolution is in Section~\ref{sec:res}. We summarise
our results in Section~\ref{sec:conc}.

\section{NGC~6240 physical characteristics}\label{sec:phys_param}

As mentioned before first spatially resolved SMBHB was a system at interacting galaxy NGC~6240
(z=0.0243 \citep{Solomon1997}) which optically classified as low-ionization nuclear emission-line
region (LINER) \citep{Veron2006}. Chandra observations confirmed the presence of two luminous hard
X-ray emission sources at south (S) and north (N) cores with a strong neutral Fe K$_{\alpha}$ line
at each \citep{Komossa2003} associated to BHs. Using the adaptive optics at Keck II telescope
\citep{Max2007,Medling2011} separation was found between the components $\sim1.5\arcsec$ which
corresponds to previous radio \citep{Beswick2001,Gallimore2004} and X-ray observations \citep{Komossa2003}.
Assuming a Hubble constant of H$_{0}=68$ km s$^{-1}$ Mpc$^{-1}$ \citep{Planck2014} at the distance of
NGC~6240 $1\arcsec$ corresponds to 500 pc, respectively the cores settle on 750 pc. \cite{Kollatschny2020}
reported a third minor component S2 which placed at $0.42\arcsec$ (210 pc) from S1 (previous S) nuclei.
Below for marking south nucleus we will use label S (S1+S2) if we talk about binary system or S1/S2 if
we talk about triple system.

Earlier CO(2-1) observations indicated maximum gas concentration between the two nuclei. It was associated
with central thin disk existence \citep{Tacconi1999}. Latest ALMA observations confirm the presence of the
nuclear molecular gas bulk  $\sim9\times10^{9}\rm M_{\odot}$ within the $\sim1\arcsec$ (500 pc) between the
two nuclei \citep{Treister2020}. Lesser part of gas concentrated within the black hole influence sphere at
north and south nuclei, $\sim7.4\times10^{8}\rm M_{\odot}$ and $\sim3.3\times10^{9}\rm M_{\odot}$ respectively.
Velocity dispersion analysis links gas presence to ongoing merging process rather than disk availability \citep{Treister2020}.

Based on X-ray luminosity \citep{Engel2010} combined black hole (N+S) mass is $4\times10^{8}\rm M_{\odot}$.
Some unobvious assumptions at this method gave uncertainty at least a factor of a few. Central part of S
nucleus inside the sphere of influence of black hole was resolved using adaptive optics \citep{Medling2011}.
Stellar dynamics gave limits for S black hole mass from $8.7\pm0.3\times10^{8}\rm M_{\odot}$ to $2.0\pm0.2\times10^{9}\rm M_{\odot}$.
Finally based on $M_{\bullet}-\sigma$ relation \cite{Kollatschny2020} found masses for north and south BHs respectively:
$M_{\bullet \rm N}=3.6\pm0.8\times10^{8}\rm M_{\odot}$, $M_{\bullet \rm S1}=7.1\pm0.8\times10^{8}\rm M_{\odot}$, $M_{\bullet \rm S2}=9.0\pm0.7\times10^{7}\rm M_{\odot}$.
Mass definitions discrepancies are attributed to system being in active merging state.

Observable NGC~6240 geometry and star formation history is shaped by first encounter and final coalescence of parent
galaxies \citep{Engel2010,Kollatschny2020}. According to Jeans model after gas fraction substractions, stellar masses
in radius 250 pc around north and in radius 320 pc around south BHs are $2.1\times10^{9}\rm M_{\odot}$ and $1.1\times10^{10}\rm M_{\odot}$
respectively. Total progenitors bulge masses from $M_{\bullet}-M_{\rm bulge}$ relation are $8.7\times10^{10}\rm M_{\odot}$ and $1.2\times10^{11}\rm M_{\odot}$
for north and south nuclei \citep{Engel2010}.

\section{Initial models' parameters}\label{sec:ini_param}

For our N-body simulations we need to convert the physical units into N-body units, so called H{\'e}non units \citep{Henon1971}.
Based on observations mentioned above we put two BHs at separation 1 kpc into two bulges with total stellar mass $1.3\times10^{11}\rm M_{\odot}$
taken from the upper limit from \cite{Engel2010}. We chose these parameters as our units of mass and distance:
\begin{equation}
\mathbf{MU}=M_{\ast}=1.3\times10^{11}{\rm M_{\odot}},
\end{equation}
\begin{equation}
\mathbf{RU}=\Delta R=1 \rm kpc.
\end{equation}
We obtained rescaled units of velocity and time in such way:
\begin{equation}
\mathbf{VU}=\Big(\frac{GM}{\Delta R}\Big)^{1/2}=747.6{\rm\;km\;s^{-1},}
\end{equation}
\begin{equation}
\mathbf{TU}=\Big(\frac{\Delta R^{3}}{GM}\Big)^{1/2}=1.3079\rm\;Myr.
\end{equation}
In new N-body units the light velocity is expressed by physical light velocity $c_{\rm phys}$ as:
\begin{equation}
c=c_{\rm phys}\Big(\frac{GM}{\Delta R}\Big)^{-1/2}=400\rm\; [NB].
\end{equation}

\begin{figure}[htbp]
\centering
\begin{minipage}{.99\linewidth}
\begin{tabular}{c}
\includegraphics[width=0.99\linewidth]{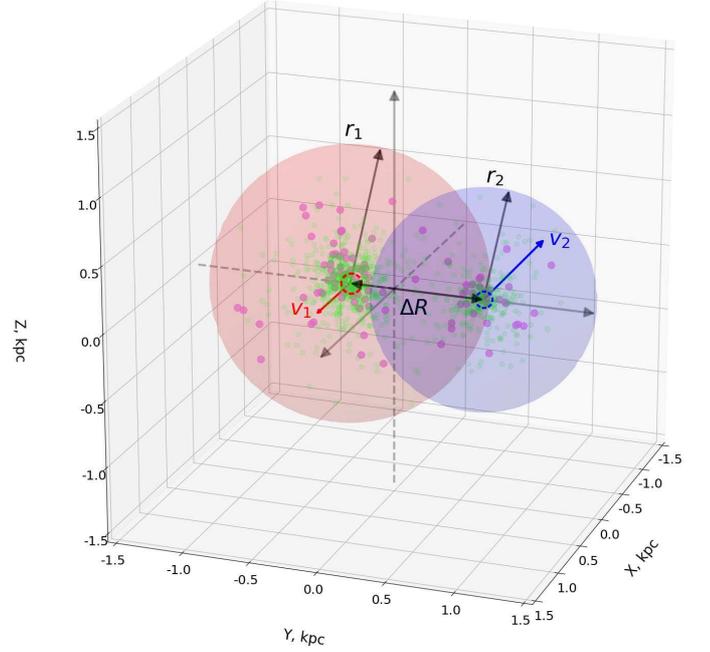}
\end{tabular}
\end{minipage}
\caption{Sketch of principal system components and initial parameters. South and north BHs are represented by red and blue
 big dashed dots respectively and placed at initial separation $\Delta R$. BHs have initial velocities $V_{1}$ and $V_{2}$,
 for south and north nuclei respectively. Bulges with sizes $r_{1}$ and $r_{2}$ (each radius equals to 5 Plummer radiuses of according bulge)
 around nuclei consist of high mass particles and low mass particles, they are represented by purple and green dots respectively.
\label{fig:model_param}}
\end{figure}

\begin{figure*}[htbp]
\centering
\begin{minipage}{.99\linewidth}
\begin{tabular}{c}
\includegraphics[width=0.99\linewidth]{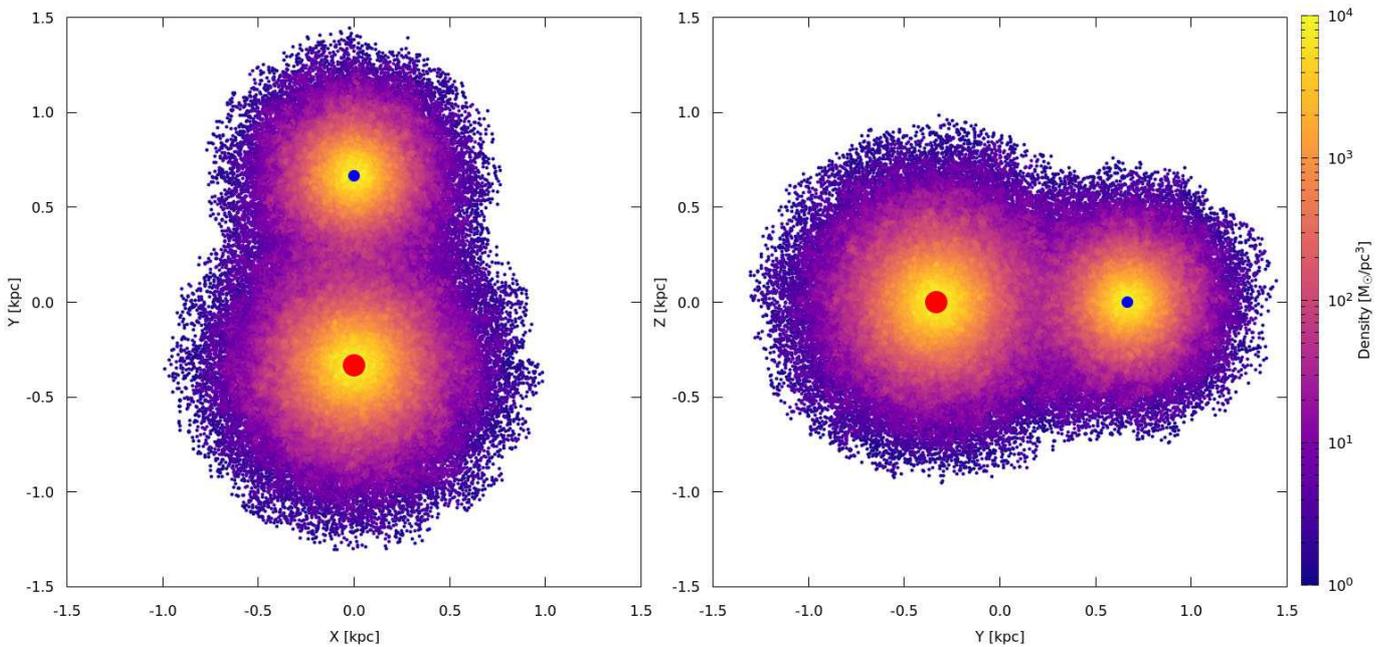}
\end{tabular}
\end{minipage}
\caption{Initial density distributions at projection planes XY and YZ for physical model A. Red and blue dots are south and north BHs respectively.
\label{fig:ini_distr}}
\end{figure*}

Our N-body models consist of two bulges with embedded central BHs (Fig.~\ref{fig:model_param}). For creating the set of \textit{physical}
models we varied mass ratio between BHs and BH to bulge. Ratio between the total BHs mass and total bulges mass was chosen
as $Q=m_{\bullet}/M_{\ast}=0.01$ (model A and C on Table~\ref{tab:models1}) and 0.02 (model B and D on Table~\ref{tab:models1}).
Mass ratio for BHs and bulge is increased in comparison to predicted value (\cite{Kormendy2013} and reference therein) due to active
coalescence of galaxy. Interacting galaxy NGC~6240 most likely represents the major merging \citep{Medling2011} and in this case we
set mass ratios between the black holes $q=m_{\bullet2}/m_{\bullet1}=0.5$ (model A and B on Table~\ref{tab:models1}) and $q=0.25$
(model C and D on Table~\ref{tab:models1}). Mass of the primary (heavier) BH was obtained in range $8.7-20.8\times10^{8}\rm M_{\odot}$
and for secondary (lighter) BH in range $2.6-8.7\times10^{8}\rm M_{\odot}$. Primary and secondary BHs, which representing South and
North nuclei, are considered special relativistic particles and located in bulges centres at separation $\Delta R$ on XY plane
with initial velocities $V_{1,2}$. The initial orbital eccentricity of the two bulges in the simple point mass approximation
was chosen as  0.5. BHs at this separation are not bound to each other.

Density distribution for each bulge is described by Plummer spheres \citep{Plummer1911}:
\begin{equation}
\rho_{1,2}(r) = \frac{3}{4\pi} \frac{M_{\ast1,2}}{R_{1,2}^{3}} \Bigl(1+\Bigl(\frac{r}{R_{1,2}}\Bigr)^{2}\Bigr)^{-\frac{5}{2}},
\end{equation}
where $M_{\ast1,2}$ is the total stellar mass and $R_{1,2}$ is Plummer radius (Fig.~\ref{fig:ini_distr}).
The same central density in each galaxy is provided by scaling the mass, size and number of particles
consistently with BH mass ratio. Fixed Plummer radius for primary bulge $R_{1}=200$ pc gave us Plummer
radius for secondary bulge $R_{2}=160$ pc for models A, B and $R_{2}=126$ pc for models C, D. After the
generation of the particles distribution we simply add to the system the two central SMBH. Due to the
very short dynamical merging scale (few Myr) of the two bulges, the initial Plummer distribution of
the particles are quickly mixed up. Separation between the centres of bulges is 1 kpc according to
the observable BH's separation.

To mimic the presence of the molecular gas cloudy and clumpy mass structure in the bulges we introduce to
our bulge particle system the multi mass prescription. Namely: 'high mass' particles (HMPs), which
represent perturbers and are associated with giant molecular clouds and/or stellar clusters, and
'low mass' particles (LMPs), which represent field particles and are associated with individual stars.
The main set of runs was done using total number of particles 500k (except model B1), where the number
fraction of HMP is 10\%. At each bulge the total HMP mass we set equal to the total LMP mass. For main
set of runs (except models B4 and B1, see details below) we use HMP mass $\sim1.3\times10^{6}\rm M_{\odot}$
and LMP mass $\sim1.4\times10^{5}\rm M_{\odot}$.

Additionally for minimizing the effect of initial numerical parameters we have constructed \textit{numerical} models
based on \textit{physical} model B (Table~\ref{tab:models1}). We use larger number of particles $N=1000$k for model
B1 (with HMP mass $\sim6.5\times10^{5}\rm M_{\odot}$ and LMP mass $\sim7\times10^{4}\rm M_{\odot}$), new
randomization for initial particles coordinates and velocities for model B2, different starting point for
$\mathcal{PN}$ terms turning on for model B3 (see Section~\ref{sec:PN_hard}), replacing HMP with LMP for
model B4 with LMP mass $\sim2.6\times10^{5}\rm M_{\odot}$.

\begin{table}
\caption{Basic N-body models.}
\label{tab:models1}
\centering
\begin{tabular}{c c c c c c c}
\hline
\hline
\multicolumn{7}{c}{\textit{Physical} models} \\
\hline
\hline
Run & $q$ & $Q$ & $m_{\bullet1}$        & $m_{\bullet2}$,       & $M_{\ast1}$,           & $M_{\ast2}$, \\
    &     &     & $10^{8}\rm M_{\odot}$ & $10^{8}\rm M_{\odot}$ & $10^{10}\rm M_{\odot}$ & $10^{10}\rm M_{\odot}$ \\
\hline
A  & 0.5  & 0.01 & 8.67  & 4.33 & \multirow{2}{*}{8.67} &  \multirow{2}{*}{4.33}   \\
B  & 0.5  & 0.02 & 17.34 & 8.66 &    &  \\
C  & 0.25 & 0.01 & 10.4  & 2.6  & \multirow{2}{*}{10.4} & \multirow{2}{*}{2.6}   \\
D  & 0.25 & 0.02 & 20.8  & 5.2  &    &  \\
\hline
\hline
\multicolumn{7}{c}{\textit{Numerical} models}\\
\hline
\hline
Run & \multicolumn{6}{c}{Characteristic property}\\
\hline
B1 & \multicolumn{6}{c}{number of particles $N=1000$k}\\
B2 & \multicolumn{6}{c}{randomisation seed}\\
B3 & \multicolumn{6}{c}{starting point for $\mathcal{PN}$}\\
B4 & \multicolumn{6}{c}{multi mass prescription}\\
\hline
\end{tabular}
\end{table}

\section{N-body code}\label{sec:code}

For our simulations we used the publicity available $\varphi$-GPU\footnote{\tt ftp://ftp.mao.kiev.ua/pub/berczik/phi-GPU/}
code with the blocked hierarchical individual time step scheme and a $4^{th}$-order Hermite integration scheme of the
equation of motions for all particles \citep{Berczik2011}. The individual timestep bin from the generalized
``Aarseth'' type criteria \citep{Makino1992,Nitadori2008}:
\begin{equation}
\Delta t_{i} = \eta_{p} \frac{A^{(1)}}{A^{(2)}} = \eta_{p} \sqrt{\frac{|\mathbf{a}||\mathbf{a}^{(2)}| + |\mathbf{\dot{a}}|^{2}}{|\mathbf{\dot{a}}||\mathbf{a}^{(3)}| + |\mathbf{a}^{(2)}|^{2}}}
\end{equation}
where $\eta_{p}$ is parameter which controls the accuracy, $\mathbf{a}^{(k)}$ is the $k^{th}$ derivative of
acceleration. In common block time step scheme the individual time steps $\Delta t_{i}$ are replaced by block
time steps $\Delta t_{i,\rm b}=(1/2)^{n}$, where $n$ satisfies:
\begin{equation*}
\Bigl(\frac{1}{2}\Bigr)^{n}\leq\Delta t_{i}<\Bigl(\frac{1}{2}\Bigr)^{n-1}
\end{equation*}
For minimization of the computational time the gravitational forces affecting the particle are divided into two types
according to Ahmad-Cohen scheme. Regular forces apply from neighbour particles and irregular forces apply from more
distinct particles, which have different time steps. It gives us the possibility to calculate irregular forces less
often than regular forces.

In case of close encounters we should carefully integrate the orbit therefore we write the acceleration for $i$
particle from $j$ particle with mass $m_{j}$ placed on separation $r_{ij}$ in form:
\begin{equation}
\mathbf {a}_{i} = - \sum\limits_{j=1,\;j\neq i}^N \frac{G m_{j}}{(r^{2}_{ij}+\varepsilon_{ij}^{2})^{3/2}} \mathbf{r_{ij}},
\end{equation}
\begin{equation}
\varepsilon_{ij}^2 = \alpha \frac{\varepsilon^{2}_{i} + \varepsilon^{2}_{j}}{2}
\end{equation}
where $\varepsilon$ is individual softening parameter for each type of particles and $\alpha$ is reducing factor.
In case of BH, HMP and LMP the softening parameters is $\varepsilon_{\rm BH}=0$, $\varepsilon_{\rm HMP}=10^{-4}$
and $\varepsilon_{\rm LMP}=10^{-5}$ respectively. When BHs interact with passing stars we reduce softening
parameter using coefficient $\alpha=10^{-4}$ for more accurate calculation of acting forces. I.e. for these
type of interaction we have the effective softening in a level of $7.1\times10^{-7}$ or $7.1\times10^{-8}$
respectively for the HMP and LMP. The individual softening in the code and also using the reducing factor
for softening allow us more accurately taking in account the BH’s dynamical evolution even up to the
$\mathcal{PN}$ merging phase.

The N-body simulations were run on five clusters: \verb"golowood" at Kyiv (Ukraine), \verb"laohu" at
Beijing (China), \verb"kepler" at Heidelberg (Germany), \verb"pizdaint" at Zurich (Switzerland), \verb"juwels"
at J\"ulich (Germany).

\section{Post-Newtonian BH and hardening}\label{sec:PN_hard}

\begin{table*}[htbp]
\caption{Characteristic times for \textit{physical} and \textit{numerical} models}
\label{tab:merge_time}
\centering
\begin{tabular}{c c r c c c c c c c c}
\hline
\hline
\multirow{2}{*}{Run} & \multirow{2}{*}{$t_{\rm b}$} & \multirow{2}{*}{$t_{\rm h}$}
                     & \multicolumn{4}{c}{$s_{\rm gw}=0.3\%s_{\rm nb}$} & \multicolumn{4}{c}{$s_{\rm gw}=3\%s_{\rm nb}$} \\
                     & & & Subrun & $t_{\rm beg}$, NB & $t_{\rm beg}$, Myr & $t_{\rm merg}$, Myr
                         & Subrun & $t_{\rm beg}$, NB & $t_{\rm beg}$, Myr & $t_{\rm merg}$, Myr \\
(1) & (2) & (3) & (4) & (5) & (6) & (7) & (8) & (9) & (10) & (11) \\
\hline
\hline
A & 5.15 & 11.12 & A.18   & 18 & 23.5 & 40.3 & A.25  & 25 & 32.7 & 46.7 \\
B & 4.38 &  9.81 & B.15   & 15 & 19.6 & 47.5 & B.23  & 23 & 30.0 & 46.5 \\
C & 9.71 & 15.04 & C.18   & 18 & 23.5 & 47.0 & C.22  & 22 & 28.8 & 52.3 \\
D & 6.58 & 11.12 & D.10   & 10 & 13.1 & 22.8 & D.12  & 12 & 15.7 & 24.7 \\
\hline
B1 & 4.15 &  9.81 & B1.11 & 11 & 14.4 & 46.7 & B1.16 & 16 & 20.1 & 39.3 \\
B2 & 4.13 & 11.12 & B2.11 & 11 & 14.4 & 44.7 & B2.16 & 16 & 20.1 & 45.3 \\
B3 & 4.30 & 11.12 & \multicolumn{7}{c}{}                         & 51.3 \\
B4 & 4.25 &  9.81 & \multicolumn{7}{c}{}                         & 31.3\\
\hline
\end{tabular}
\tablefoot{Columns: (1) runs' names for \textit{physical} A, B, C, D and \textit{numerical} B1, B2, B3, B4 models;
(2) bounding time for central SMHBs;
(3) hardening time for central SMHBs;
(4),(8) $\mathcal{PN}$ runs' names;
(5),(9) beginning time for $\mathcal{PN}$ runs in H\'enon units;
(6),(10) beginning time for $\mathcal{PN}$ runs in Myr;
(7),(11) merging time in Myr.
Columns (4)-(7) correspond to $s_{\rm gw}=0.3\%s_{\rm nb}$, columns (8)-(11) correspond to $s_{\rm gw}=3\%s_{\rm nb}$.
For models B3, B4 we turned on $\mathcal{PN}$ terms from start point.
}
\end{table*}

As mentioned above BHs are embedded as special massive relativistic particles located at the centres each bulges.
The equation of motion was written via accelerations expressed through positions and velocities in $\mathcal{PN}$
formalism. $\mathcal{PN}$ theory is an approximate version of general relativity therefore in $\mathcal{PN}$
terms up to 3.5$\mathcal{PN}$ equation of motion schematically can be written in form
\begin{equation}
\mathbf{a} = \mathbf{a}_{\mathcal{PN}} + \frac{1}{c^{2}}\mathbf{a}_{1\mathcal{PN}} + \frac{1}{c^{4}}\mathbf{a}_{2\mathcal{PN}} + \frac{1}{c^{5}}\mathbf{a}_{2.5\mathcal{PN}} +
             \frac{1}{c^{6}}\mathbf{a}_{3\mathcal{PN}} + \frac{1}{c^{7}}\mathbf{a}_{3.5\mathcal{PN}}
\end{equation}
where $a_{\mathcal{PN}}$ is the classical Newtonian acceleration, $a_{1\mathcal{PN},2\mathcal{PN},3\mathcal{PN}}$
correspond to energy conservation,  $a_{2.5\mathcal{PN},3.5\mathcal{PN}}$ correspond to gravitational wave emission.
Acceleration in the center-of-mass frame is
\begin{equation}
a = \frac{d\mathbf{V}}{dt}= - \frac{GM}{\Delta R^{2}} [(1+A)\mathbf{n}+B\mathbf{V}],
\end{equation}
where $\mathbf{n}$ is normalized position vector, $\mathbf{V}$ is relative velocity \citep{Blanchet2006}.
Coefficients A and B (see equations (182), (183), (185), (186) at \cite{Blanchet2006} and references therein)
are the complex functions of masses, velocities and separations.

Hard binary evolution can be split into two distinct regimes: classical and relativistic. Driven by the
stellar-dynamical effects hardening in classical regime is constant during the long period:
\begin{equation}
s_{\rm nb} =\frac{d(1/a)}{dt} \approx \rm const.
\end{equation}
At the next phase gravitational wave emission starts to be dominating dissipation force and hardening drastically
decreases. After relativistic and classical hardenings are equal, in case of just 2.5$\mathcal{PN}$ corrections,
average eccentricity and semimajor axis evolution is described as \citep{Peters1963,Peters1964a,Peters1964b}:
\begin{equation}
\Bigl\langle\frac{da}{dt}\Bigr\rangle_{\rm gw} = - \frac{64}{5}
\frac{G^{3} m_{\bullet1} m_{\bullet2} (m_{\bullet1}+m_{\bullet2})}{a^{3} c^{5} (1-e^{2})^{7/2}}
\Bigl( 1 + \frac{73}{24}e^{2}+ \frac{37}{96}e^{4} \Bigr)
\end{equation}
\begin{equation}
\Bigl\langle\frac{de}{dt}\Bigr\rangle_{\rm gw} = - \frac{304}{15} e
\frac{G^{3} m_{\bullet1} m_{\bullet2} (m_{\bullet1}+m_{\bullet2})}{a^{4} c^{5} (1-e^{2})^{5/2}}
\Bigl( 1 + \frac{121}{304}e^{2} \Bigr)
\end{equation}
One can obtain the binary hardening during relativistic regime caused by gravitation wave emission \citep{Khan2012}
\begin{equation}
\label{eq:gw_hard}
s_{\rm gw} = \frac{64}{5}
\frac{G^{3} m_{\bullet1} m_{\bullet2} (m_{\bullet1}+m_{\bullet2})}{a^{5} c^{5} (1-e^{2})^{7/2}}
\Bigl( 1 + \frac{73}{24}e^{2}+ \frac{37}{96}e^{4} \Bigr)
\end{equation}

\section{Results and discussion}\label{sec:res}

The SMBHB evolution consists of three main time scales: period of time before
gravitational bounding, the hardening time frame and a the time interval with
strong GWs radiation emitting. The first two periods are described by the
classical Newtonian dynamics. During the second phase we have a almost constant
N-body hardening rate $s_{\rm nb}$ in short timescale. The last time
interval is a relativistic mode with a combined classical and relativistic hardening
rates $s_{\rm tot} = s_{\rm nb} + s_{\rm gw}$, where the latter term is dominating.
Due to this we follows the next strategy.

As mentioned before, our SMBHB model at initial separation of 1~kpc is not a bound system.
At first step we found the time of bounding our model binary system in a pure N-body runs  and
fitted the inverse semimajor axis $1/a$ with the linear function. The slope of this linear function is
our constant N-body hardening. At the second step we obtain the theoretical relativistic
hardening directly from~(\ref{eq:gw_hard}). At the last third step we compare these hardening's and
choose the starting points to turn on the $\mathcal{PN}$ terms using two criteria $s_{\rm gw}=0.3\%s_{\rm nb}$ and
$s_{\rm gw}=3\%s_{\rm nb}$. In totals for models A, B, C, D, B1, B2 we have one N-body run without $\mathcal{PN}$ terms
(6 runs total) and two $\mathcal{PN}$ runs for each of Newtonian runs with different $\mathcal{PN}$ terms starting points
$t_{\rm beg}$ (12 runs total) (Table~\ref{tab:merge_time}). Relativistic sub-runs contain a number which presents
a time when we turning on our $\mathcal{PN}$ terms. For models B3 and B4 we turn on $\mathcal{PN}$ terms directly
from $t_{\rm beg}=0.0$. So, in total we have 20 individual runs.

\begin{figure*}
\centering
\includegraphics[angle=-90,width=0.9\linewidth]{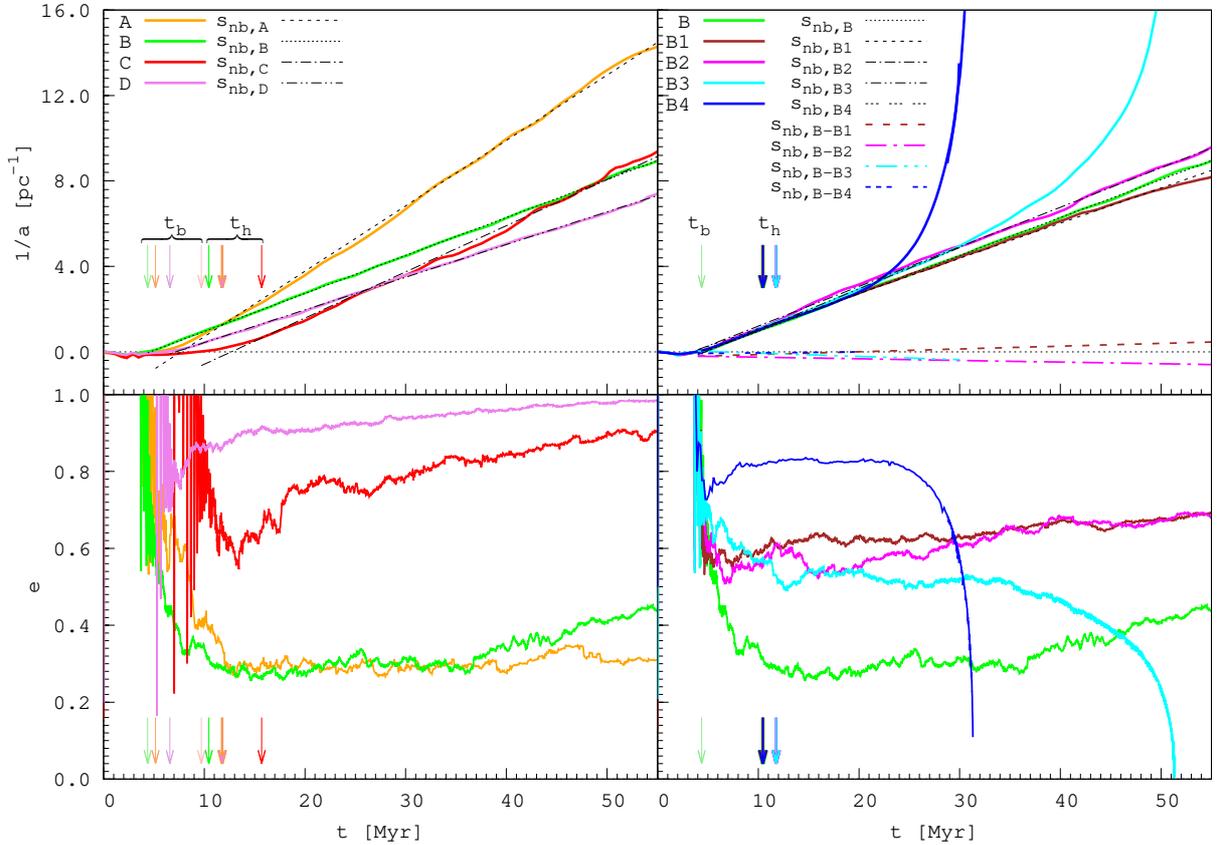}
\caption{From top to bottom: evolution of inverse semimajor axis eccentricity in pure
N-body regime for \textit{physical} models A, B, C, D (left) and \textit{numerical} models B1-4
(right) based on \textit{physical} model B. Of note, for \textit{numerical} models B3 and B4
we turned on $\mathcal{PN}$ terms from the start of simulations.
For \textit{numerical} models residuals between the model B and models B1-4 are shown as color
dashed lines. Color arrows mark time when each system getting bound $t_{\rm b}$ (pale color)
and  hard $t_{\rm h}$ (intensive color). Colors chosen by main color scheme for each model.
\label{fig:hard-e-NB}}
\end{figure*}

\textit{Physical} and \textit{numerical} models in pure N-body regime (Fig.~\ref{fig:hard-e-NB})
show quick BHs bounding from beginning of simulations at time $t<10$ Myr (Table~\ref{tab:merge_time})
with maximum value for model C. Binary starts to be hard at the time $t_{\rm h}<12$ Myr with
longest time in model C due to the longest bound time. This time estimation has the error $\pm0.66$
Myr because of time interval between the snapshots of the runs. Shrinking of the orbits in
classical regime happens due to the interaction with surrounding stars. Semi-major axis is
constantly decreasing as a linear function from gravitational bounding up to almost the merging time.
With time the amount of particles which interact with binary drops, so called loss-cone depletion,
and a hardening can gets more flat. For our simulations the $\mathcal{PN}$ relativistic terms
start to working before this happens. So, in our cases the hardening does never stall. In all of
our different \textit{physical} models, as a primary BH mass increases, we see that the model
inverse semimajor axis does not depend on the binary orbital parameters.

The different \textit{numerical} models show a good agreement of hardening rates.
It shows independence of hardening rates from number of particles (in agreement with
\cite{Berczik2006,Berentzen2009}), initial randomization (in agreement with
\textcolor{blue}{Kompaniiets et al. 2021, in preparation}), starting point
of $\mathcal{PN}$ terms.
For \textit{physical} models C and D bounding pairs eccentricity (0.6-0.9) is higher
than for the models A and B (0.3-0.4), but since in our N-body simulations the
eccentricity is more or a less random, due to the stochastic nature of this process
\citep{Wang2014, Quinlan1996}.
For \textit{numerical} models with non-physical differences we see eccentricity at
a level of 0.6-0.8. Eccentricity slowly grows with time for both \textit{physical}
and \textit{numerical} models \citep{Preto2011}.

\begin{figure*}
\centering
\begin{minipage}{.48\linewidth}
\includegraphics[width=0.99\linewidth]{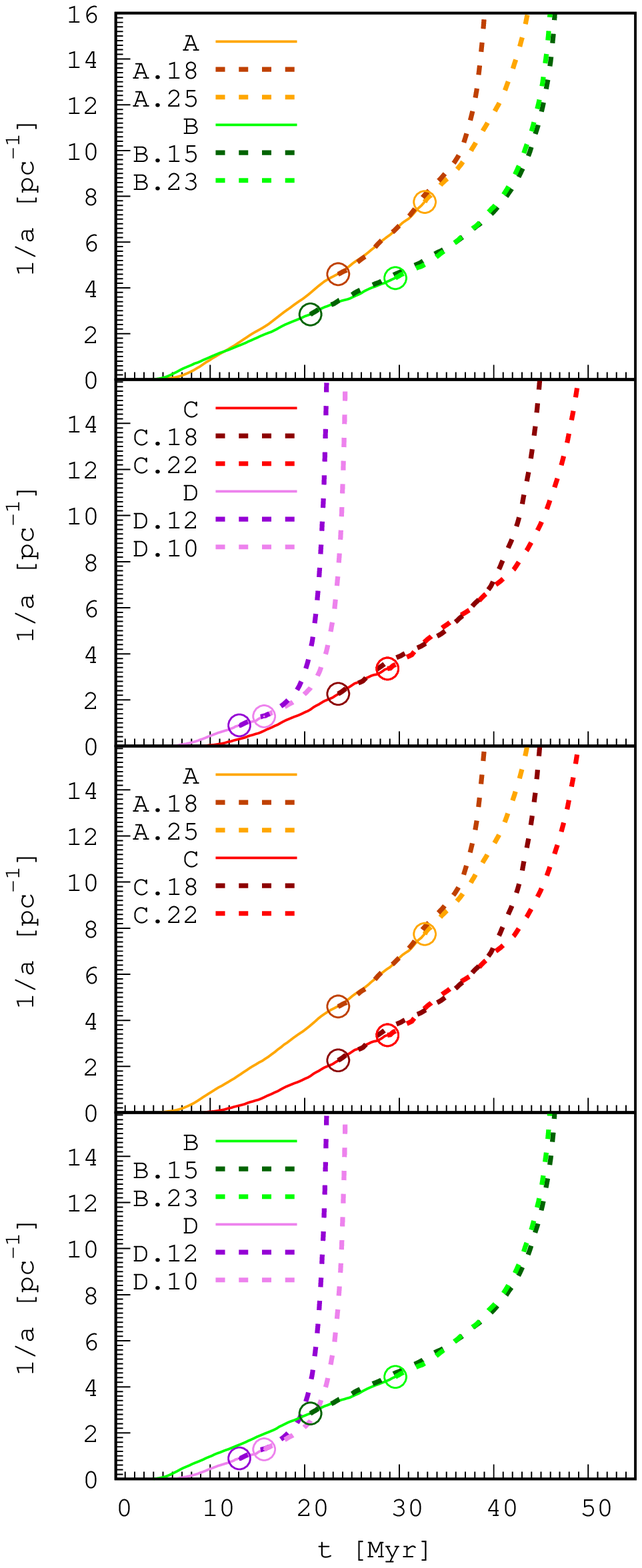}
\end{minipage}
\centering
\begin{minipage}{.48\linewidth}
\includegraphics[width=0.99\linewidth]{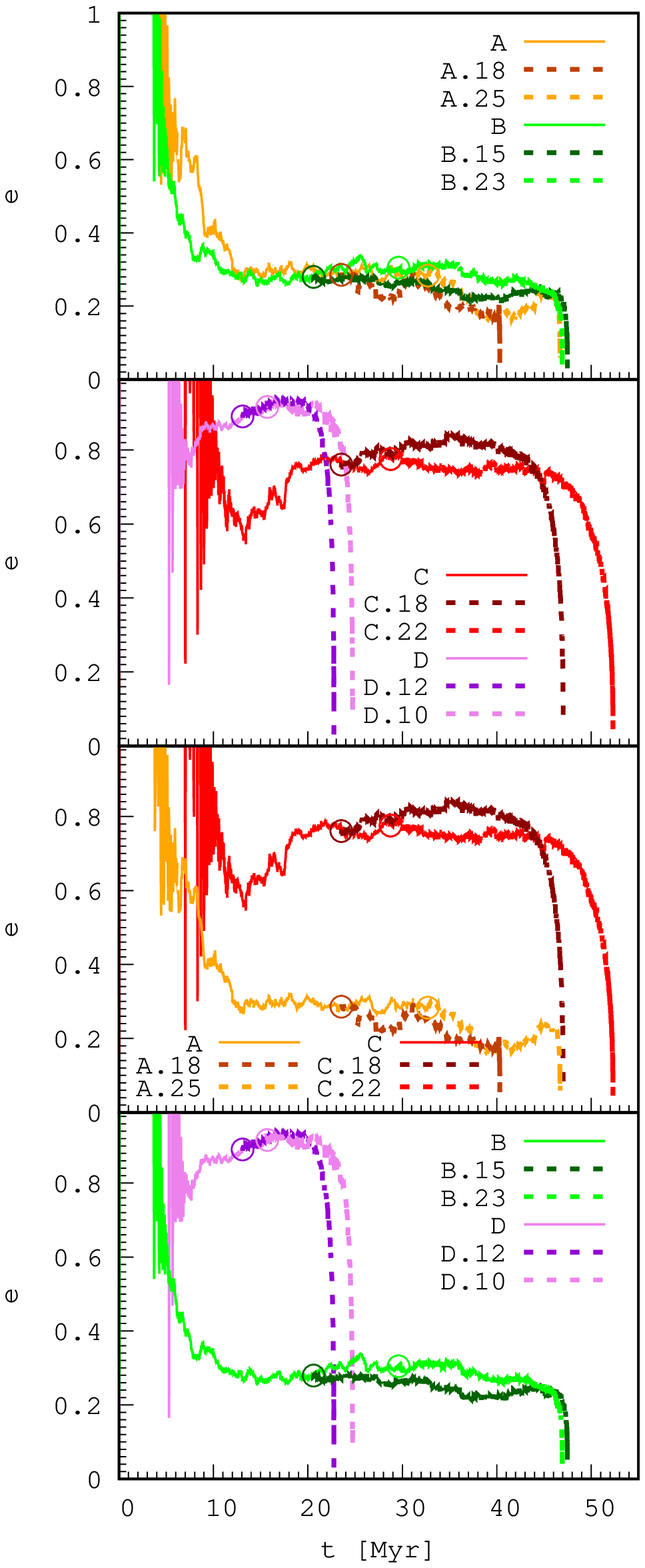}
\end{minipage}
\caption{Inverse semimajor axis (left) and eccentricity (right) evolution for \textit{physical}
models A, B, C, D composed of N-body and $\mathcal{PN}$ runs. Start point for turning on
$\mathcal{PN}$ terms from Table~\ref{tab:merge_time} noted by circles. Evolution for pure N-body
regime are cut at time when latter $\mathcal{PN}$ runs start.
\label{fig:ABCD-a-e}}
\end{figure*}

\begin{figure*}
\centering
\begin{minipage}{.48\linewidth}
\includegraphics[width=0.99\linewidth]{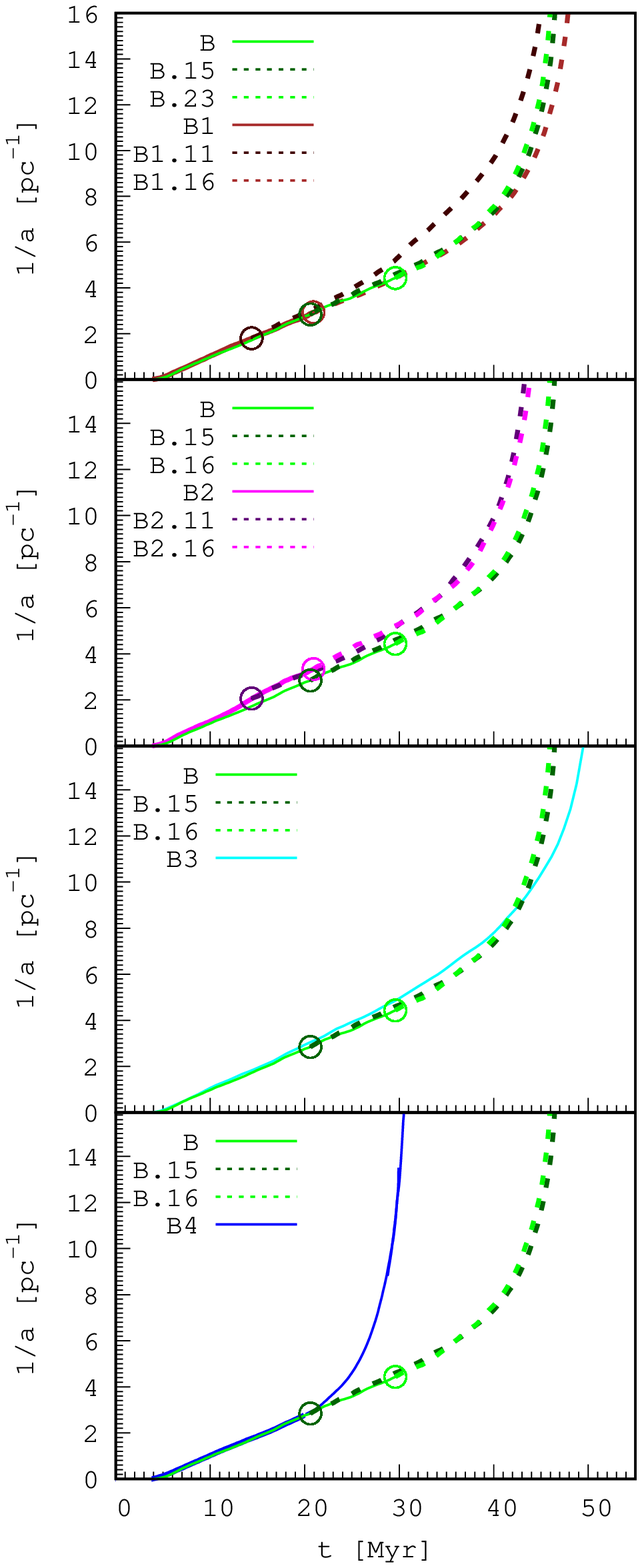}
\end{minipage}
\centering
\begin{minipage}{.48\linewidth}
\includegraphics[width=0.99\linewidth]{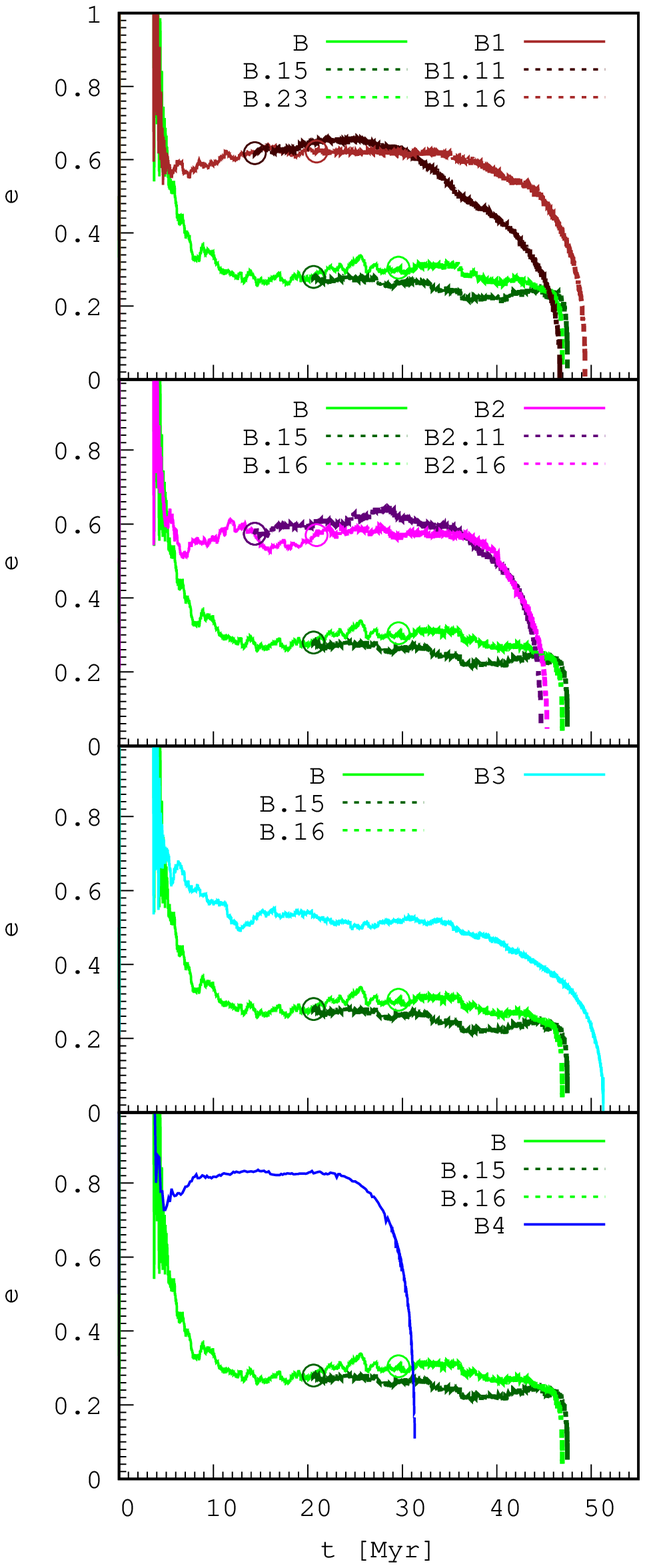}
\end{minipage}
\caption{Inverse semimajor axis (left) and eccentricity (right) evolution for \textit{numerical}
models B1-4 composed of N-body and $\mathcal{PN}$ runs. Start point for turning on $\mathcal{PN}$
terms from Table~\ref{tab:merge_time} noted by circles. Evolution for pure N-body regime are cut
at time when latter $\mathcal{PN}$ runs start.
\label{fig:BB1B2B3B4-a-e}}
\end{figure*}

Evolution with $\mathcal{PN}$ terms turned on shows a binary merging due to GW emission
(Fig.~\ref{fig:ABCD-a-e}, Fig.~\ref{fig:BB1B2B3B4-a-e}). We assume that the merging itself
happens when the separation between the components is less than 4 Schwarzschild radius that
equal 5.2 mpc for the models with $Q=0.01$ and 10.4 mpc for the models with $Q=0.02$.
Model D with biggest mass of secondary BH has the shortest run and a model C with lowest
mass of primary BH has the longest run. Inverse semimajor axis shows steady increase in
classical regime and rapid rise at time when the GWs emission is dominant
in dissipative force (Fig.~\ref{fig:ABCD-a-e}, Fig.~\ref{fig:BB1B2B3B4-a-e} left panels).
Merging time for \textit{physical} models is on ranges from 22 Myr to 53 Myr
(Table~\ref{tab:merge_time}, also see \cite{Sobolenko2016}). Eccentricity shows the same
merging behavior in classical and relativistic regimes (Fig.~\ref{fig:ABCD-a-e},
Fig.~\ref{fig:BB1B2B3B4-a-e} right panels): after the binary is formed, eccentricity is
almost constant in a classical regime and approaching to 0 (circular orbit) in a relativistic
regime. Interestingly, the fastest-merging model D has the highest mass of primary BH and
eccentricity around 0.9.

Compared to \textit{physical} model B, models
B1 with different number of particles and B2 with different initial randomization of
particles positions and velocities do not show significant differences (also see
\textcolor{blue}{Kompaniiets et al. 2021, in preparation}). The choice of
starting point $t_{\rm beg}$ does not affect merging time notably even for model B3
with $\mathcal{PN}$ terms turned on from the beginning. For models B1 and B2 the eccentricity
is at the level of 0.6, for model B3 the eccentricity is a little bit less, at the level
0.5-0.6, for model B4 the eccentricity is higher and it is at the level 0.8.  Model B4
with just a low mass particles merges even more quickly in time $\sim$30 Myr.  Total
differences in merging time for all of our numerical models is less then 10\%
(Table~\ref{tab:merge_time}). Obtained merging time is larger than in research
\textcolor{blue}{Kompaniiets et al. 2021, in preparation} due to different initial physical
parameters, such as total bulge mass and separation.

For each run the merging time for both \textit{physical} and \textit{numerical} models does not
significantly change due to the variations of the $\mathcal{PN}$ terms starting
points $t_{\rm beg}$. At the same time our technique of using two series
(with 0.3\% and 3\% GW hardening compared to the N-body dynamical hardening) for
hardening, gives us a quite good computational time savings (Table~\ref{tab:comp_time}).

\begin{table}
\caption{Physical computational time at clusters.}
\label{tab:comp_time}
\centering
\resizebox{\columnwidth}{!}
{%
\begin{tabular}{c c l c}
\hline
\hline
Run                        & Node number & GPU card        & Time      \\
\hline
\hline
B3(NB+$\mathcal{PN}$)      & 16 nodes    & Tesla K20       & 27.3 days \\
B(NB)+B.23($\mathcal{PN}$) & 10 nodes    & Tesla K20       & 2.6 days  \\
                           &             & GeForce GTX 660 &           \\
\hline
\end{tabular}%
}
\end{table}

\begin{figure*}
\centering
\begin{minipage}{.48\linewidth}
\includegraphics[width=0.99\linewidth]{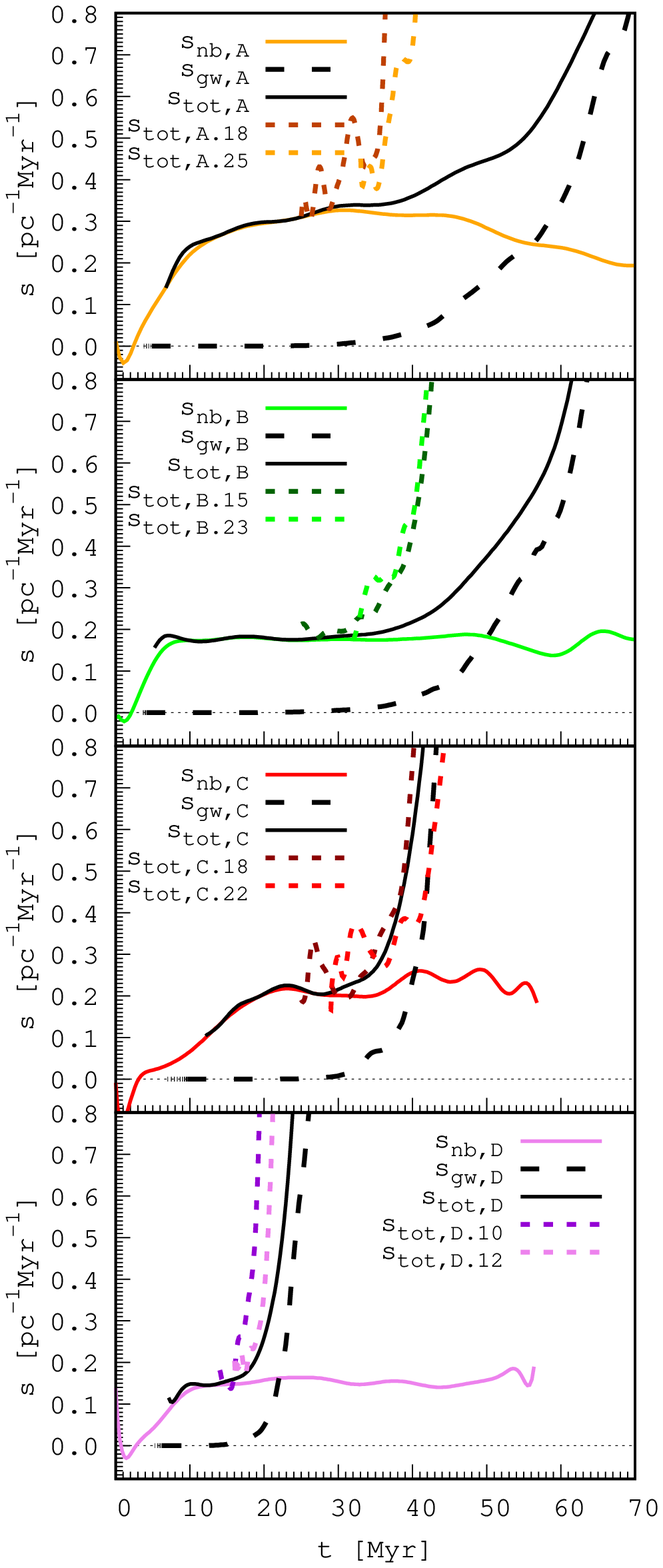}
\end{minipage}
\centering
\begin{minipage}{.48\linewidth}
\includegraphics[width=0.99\linewidth]{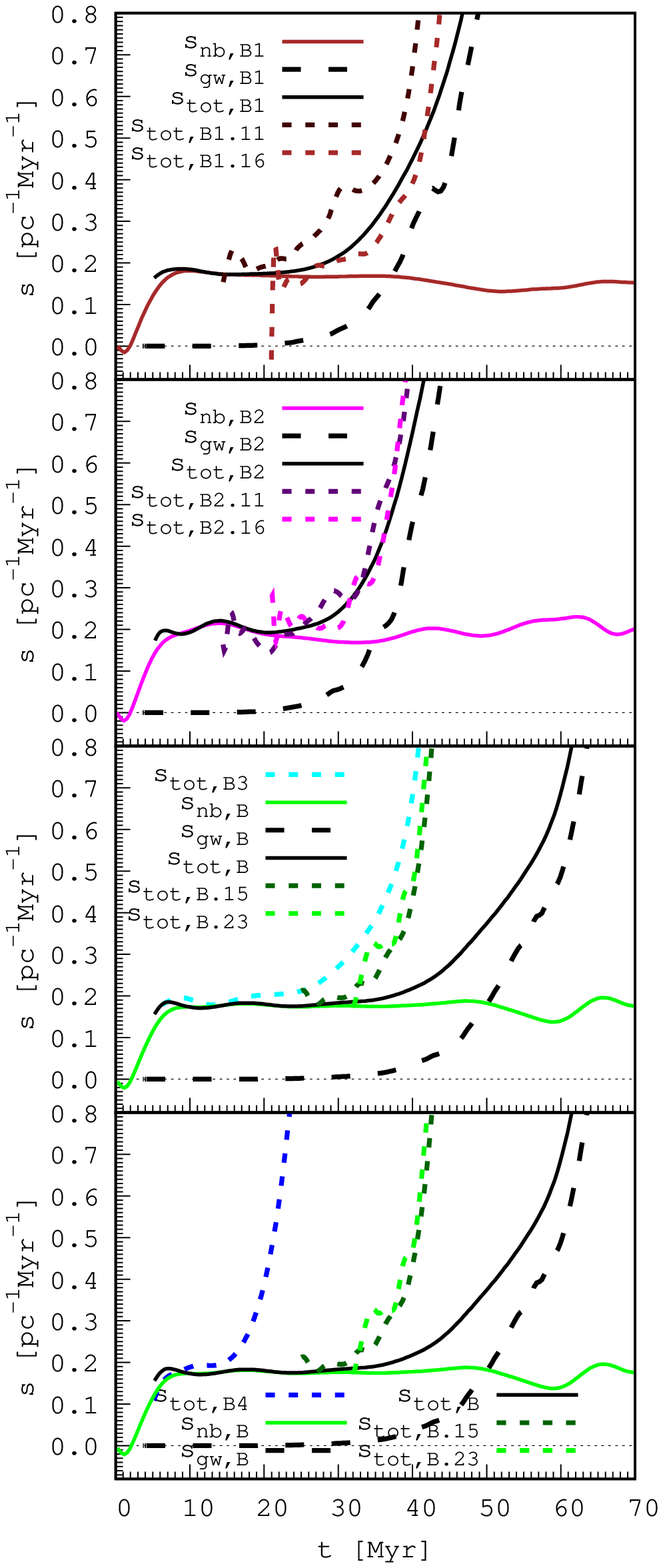}
\end{minipage}
\caption{Hardening comparison for \textit{physical} A, B, C, D (left) and \textit{numerical}
B1-4 (right) models from N-body simulations and theoretical relativistic hardening from~(\ref{eq:gw_hard})
\label{fig:s_gw}}
\end{figure*}

To check the accuracy of choosing the proper starting point we aligned the theoretical models
hardening in relativistic regime (Fig.~\ref{fig:s_gw}). The model's relativistic hardening
for \textit{physical} runs of major merging A and B (with mean eccentricity at level of 0.3)
shows a earlier decrease than the simple theoretical value. The merging time differences for
model A is 25 Myr and for model B is around 20 Myr. For \textit{physical} models C and D
(with mean eccentricity is $>0.8$) the modeling relativistic hardening is at the same order
as a theoretical value. For \textit{numerical} models (with mean eccentricity $>0.5$) the
difference between the theoretical model values is not so high and the best matching is
observed in the models B3 and B4, where we turn on $\mathcal{PN}$ terms from the beginning.
\textit{Numerical} models B, B2, B3 have a same physical base, i.e. mass of BHs, similar N-body
hardening but differs in binary eccentricity. According restricted number of our models,
we did not find dependence of eccentricity and parameters of our systems. As expected from previous
results the binary can form system with any initial eccentricity and statistically was found the
trend to form more circular binary for steeper cusps \citep{Khan2018}. Also \citep{Nasim2020}
demonstrated the tendency of decreasing the eccentricity scattering with number of particles.

\section{Conclusions}\label{sec:conc}

We have presented the direct N-body modeling with $\mathcal{PN}$ corrections up to 3.5$\mathcal{PN}$
for a central SMBHB in interacting galaxy NGC~6240. Creating the models set with a varying physical
and numerical parameters gave us the limitations of major merging times for central binary.
In a range of our parameters, we did not find any significant correlation between merging time
and BHs mass or BH to bulge mass ratios. We understand that the set of initial physical conditions,
generally speaking, can strongly effect of our merging time estimations. Also mass segregation
can strongly affect a merging time (see \cite{Gualandris2012,Khan2018}), but to discuss this issue
in context of our models with mixed mass prescription the further detail research is needed.
Varying numerical parameters (randomization, number of particles, $\mathcal{PN}$ starting point) does
not strongly affect of the merging time for BHs. The estimated time for a merging in NGC~6240 galaxy
with the different physical parameters is less than $\sim$55 Myr. But this concrete number are
valid only for our set of combination of initial mass ratios. In current case we limited our mass
ratio ranges to the galaxy NGC 6240. Our numerical technique for turning on the relativistic $\mathcal{PN}$
forces not from the beginning of the run gave us the significant computational time savings. Further
detailed research of rare dual/multiple BHs in dense stellar environment (based on observations data)
can clarify the dynamical co-evolution of central BHs and their host-galaxies.

\begin{acknowledgements}
We were partly supported by the Deutsche Forschungsgemeinschaft (DFG, German Research Foundation)
Project-ID~138713538, SFB~881 ("The Milky Way System") and by the Volkswagen Foundation under the
Trilateral Partnerships grant No.~97778.

We acknowledge partial support by the Strategic Priority Research Program (Pilot B) Multi-wavelength
gravitational wave universe of the Chinese Academy of Sciences (No.~XDB23040100).

RS acknowledges PKING (PKU-KIAA Innovation NSFC Group, gravitational astrophysics part, NSFC grant 11721303).

The work of PB and MS was supported under the special program of the NRF of Ukraine "Leading and Young
Scientists Research Support" - "Astrophysical Relativistic Galactic Objects (ARGO): life cycle of active
nucleus", No.~2020.02/0346.

MS acknowledges support by the National Academy of Sciences of Ukraine under the Research Laboratory
Grant for young scientists No.~0120U100148 and Fellowship of the National Academy of Science of Ukraine
for young scientists 2020-2022.

PB acknowledges support by the Chinese Academy of Sciences (CAS) through the Silk Road Project at NAOC,
the President’s International Fellowship (PIFI) for Visiting Scientists program of CAS and the National
Science Foundation of China (NSFC) under grant No.~11673032.

The authors gratefully acknowledge the Gauss Centre for Supercomputing (GSC) e.V. (www.gauss-centre.eu)
for funding this project by providing computing time through the John von Neumann Institute for
Computing (NIC) on the GCS Supercomputers JURECA and JUWELS at J\"ulich Supercomputing Centre (JSC).

\end{acknowledgements}

\bibliographystyle{aa}
\bibliography{ngc6240-aa}

\end{document}